# Metallo-Dielectric Multilayer Structure for Lactose Malabsorption Diagnosis through $H_2$-Breath-Test

N. Cioffi, D. de Ceglia, M. De Sario, A. D'Orazio, V. Petruzzelli, F. Prudenzano, M. Scalora, S.Trevisi, M. A. Vincenti

*Abstract*—A metallo-dielectric multilayer structure is proposed as a novel approach to the analysis of lactose malabsorption. When lactose intolerance occurs, the bacterial overgrowth in the intestine causes an increased spontaneous emission of $H_2$ in the human breath. By monitoring the changes in the optical properties of a multilayer palladium-polymeric structure, one is able to detect the patient's disease and the level of lactose malabsorption with high sensitivity and rapid response.

THE lowest layer of the terrestrial atmosphere, the troposphere, contains about 0.575ppm of hydrogen ($H_2$). In contrast, the breath exhaled by healthy people preferentially holds 20ppm or more of $H_2$. The increased hydrogen emission occurs primarily because part of ingested carbohydrates and proteins is not absorbed or digested, and is instead fermented to form hydrogen by bacteria settling either in the colon or, especially in the case of a bacterial overgrowth, in the small intestine. Part of this hydrogen is then dissolved in mucosal blood, transported to the lungs within a few minutes, and finally appears in the breath [1]. The most common cause of lactose malabsorption and lactose intolerance is a natural decrease in intestinal lactase levels, known as primary adult hypolactasia or lactase non-persistence [2]. The lactose intolerance can be hard to diagnose based on symptoms alone because the symptoms associated with this disorder are also common in other conditions, for example, irritable bowel syndrome. The most common test used to diagnose lactose malabsorption is the lactose hydrogen breath test. During this test, the patient ingests a large dose of lactose, after which hydrogen levels in the breath are measured at regular intervals of 30 or 40 minutes. A rise in hydrogen breath levels (>20ppm over the baseline) theoretically indicates increased gas production due to malabsorbed lactose reaching the large intestine. Hence, by plotting the $H_2$-concentration in the breath as a function of time after ingestion of the lactose solution, one is able to reveal a malabsorption condition. However, a positive response in the $H_2$ breath test that diagnoses lactose malabsorption may or may not mean the occurrence of lactose intolerance [3-5].

In order to realize a highly sensitive sensor for a $H_2$ breath test, we underline the importance of the selectivity in hydrogen sensing. This issue arises from the fact that in the majority of field applications, as well as in the exhaled breath, hydrogen must be detected in the presence of other gases, i.e. oxygen and nitrogen in the air. Moreover, the breath of healthy people contains nitrogen, oxygen, carbon dioxide, water vapor and inert gases. The remainder of human breath (<0.000001%) is a mixture of as many as 500 different compounds [6]. However, pathologic conditions could alter the breath composition introducing other gases and substances that can modify the sensor response. As a consequence, the presence of some traps for organic species is essential to preserve the detection mechanism from external damaging agents. A large number of materials and devices operating on different principles has been exploited for hydrogen sensing [7], such as a modified metal-oxide resistive gas sensor [8,9], pellistor [10] and ChemFET type [11] devices. Furthermore, many transition metals of the group VIII such as nickel, palladium and platinum spontaneously absorb hydrogen. This absorption causes a change of electrical and optical properties of the above mentioned metals. Palladium has a remarkable ability to absorb hydrogen. In addition, the formation of palladium hydride results in increased electrical resistance, and in decreased dielectric constant [12]. Several configurations for hydrogen sensors based on the properties of palladium have been proposed. It has been demonstrated that thin palladium films can be deposited on an optical fiber to detect hydrogen [13, 14]. The measurement of hydrogen concentrations is also possible by monitoring the reflectivity of a thin palladium film deposited on the tip of an optical fiber

D. de Ceglia, M. De Sario, A. D'Orazio, V. Pertruzzelli, S. Trevisi, M. A. Vincenti are with the Dipartimento di Elettrotecnica ed Elettronica, Politecnico di Bari, Via Orabona 4, 70125, Bari, Italy (e-mail: vincenti@ deemail.poliba.it).

F. Prudenzano is with Dipartimento di Ingegneria dell'Ambiente e per lo Sviluppo Sostenibile, Politecnico di Bari, Taranto, Via del Turismo 8, 74100, Italy.

M. Scalora is with the Charles M. Bowden Research Center, AMSRD-AMR-WS-ST, Redstone Arsenal, AL 35898-5000, USA

N.Cioffi is with Università degli Studi di Bari, Dipartimento di Chimica, Via Orabona 4, 70125, Italy.

[15], or by monitoring the optical transmittance of a thin palladium film grown on a glass substrate [16]. Finally, more sensitive but also more complex measurements could be performed by monitoring optical changes in the surface plasmon resonance, [17] or interference phenomena [18].

In this paper we propose a novel approach to the detection and inspection of hydrogen concentrations. Our analysis yields a new device useful for $H_2$-breath-test analysis. A palladium-dielectric multilayer structure reacts with small quantities of hydrogen in the human breath, leading to reversible changes in the transmission spectrum. These changes involve certain wavelengths inside the spectrum, and depend on the capability of palladium films to absorb hydrogen molecules, leading to changes in the dielectric constant.

## I. THE PALLADIUM-HYDROGEN SYSTEM

Hydrogen is a combustible substance with a lower explosive limit of 4.65% at room temperature and pressure, and since it is the smallest of the elements, one of the main barriers to overcome is the detection of small quantities of hydrogen. Due to its combustible nature, optical monitoring of hydrogen is considered the most appropriate method owing to its inherent safe nature when compared with techniques that require electrical measurements [12]. Moreover, a lot of transition metals such as nickel, palladium and platinum absorb hydrogen spontaneously, changing their optical properties as a function of hydrogen concentration. At the end of 19$^{th}$ century, it was demonstrated that palladium shows larger absorption of hydrogen compared to iron or platinum or other transition metals, and that it is possible to store a high quantity of hydrogen in a Pd lattice [19]. Free electrons states of Pd are filled by free electrons of the absorbed hydrogen, modifying the Fermi level and changing reversibly the electrical [20] and optical [21] properties of palladium. This causes the decrease of both the real and the imaginary parts of the complex permittivity of palladium [22]. It was also demonstrated that the complex permittivity of palladium hydride films $\varepsilon_{PdH}$ could be expressed as:

$$\varepsilon_{PdH} = h \times \varepsilon_{Pd} \qquad (1)$$

where $\varepsilon_{Pd}$ is the complex permittivity of bare palladium and $h$ is a nonlinear function having values between 0 and 1, that describes the dispersive behavior of Pd as a function of hydrogen concentration. Values of $h$ have been calculated and measured for a wide range of hydrogen concentrations, and they remain approximately constant if the metal thickness is changed [13]. It was also demonstrated that the Drude model could describe the hydride palladium optical properties as well, when the hydrogen concentration is increased [13,23]. Hence, to take into account the dispersion properties of bare Pd we use the well known Drude model, expressed by the following relation:

$$\varepsilon_{Pd} = \varepsilon_{\infty} - \frac{(\omega_p \tau)^2}{(1 + (\omega \tau)^2)} \qquad (2)$$

where $\varepsilon_{\infty}$ is the relative permittivity at high frequencies, $\omega_p$ is the plasma frequency, $\tau$ is the relaxation time and $\omega$ is the angular frequency. In order to express the Drude parameters as a function of hydrogen concentration, the $\varepsilon$ of the Pd thin film has been measured using different substrates, and by comparing the results in terms of transmittance and reflectance [24]. Once the nonlinear function $h$ (see eq.1) is evaluated, the Drude parameters and the dielectric constant are calculated for all the hydrogen concentrations under investigation. By interpolating $\omega_p$, $\tau$ and $\varepsilon_{\infty}$, the real (Fig. 1a) and the imaginary parts (Fig. 1b) of the dielectric constant for hydride palladium are uniquely determined. How a palladium hydride condition occurs? The Palladium hydride is formed when palladium is exposed to hydrogen, and an efficient dissociation rate occurs when hydrogen atoms are in contact with the palladium surface. The palladium hydride film has different mechanical, electrical, and optical properties than those of bare palladium. Moreover, hydride palladium exhibits a different behavior depending on the value of the hydrogen concentration. The hydride palladium could be characterized by two different phases, the α phase and the β phase. The reversible α phase occurs at low hydrogen concentrations, and in the absence of hydrogen Pd is considered in the α phase as well. If the hydrogen concentration is increased, Pd is transformed into the β phase. The transition point not only depends on the hydrogen concentration, but also on the film thickness and temperature. The β phase is not reversible and introduces an hysteresis in the optical and mechanical parameters of the palladium film. The transition from α to β phases for a thin palladium film (10-100nm) occurs at wider hydrogen concentration values (more than 4%) [25]. For hydrogen sensing applications it is important to avoid the Palladium phase transition during the experiments. A $H_2$-breath-test is based on the detection of small quantities of hydrogen, ranging from 20ppm for healthy conditions, to 80ppm or more for pathologic conditions. In the whole range of our investigation it is assumed that the palladium remains in the reversible α phase, in order to be able to repeat the measurements with the same device without any hysteresis phenomena, or other structural variations. The desorption of the hydrogen can be operated by using a pure nitrogen atmosphere.

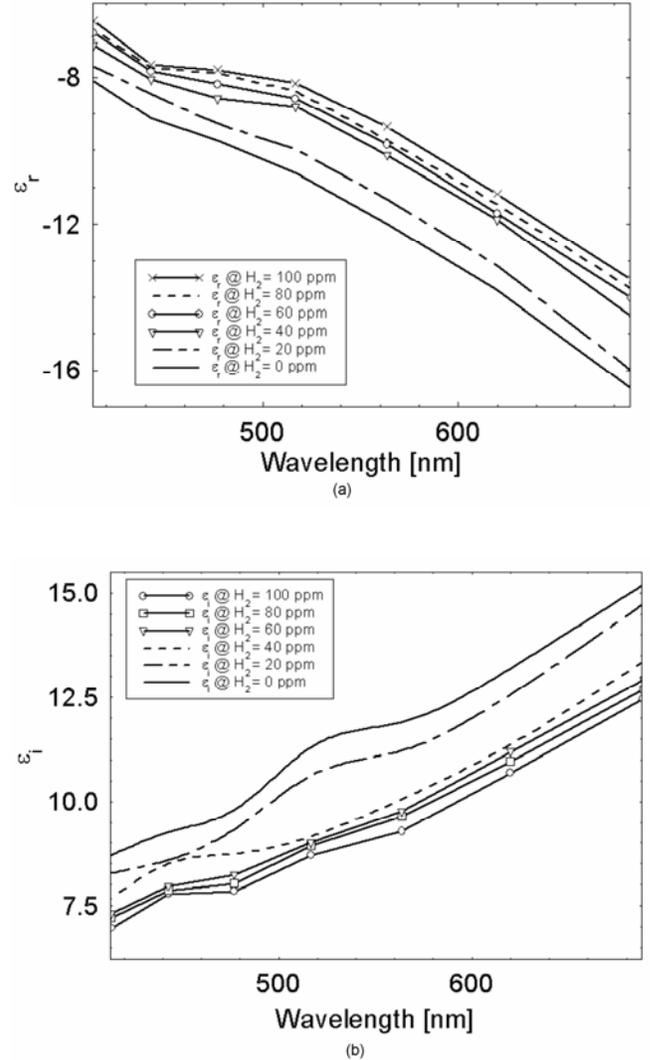

Fig.1: (a) real part and (b) imaginary part of the dielectric constant against the hydrogen concentration in air. Hydrogen concentration values are typical of hypolactasia pathologic condition.

## II. METALLO-DIELECTRIC STRUCTURE THEORY

In recent years it has been demonstrated that it is possible to realize one-dimensional photonic band gap (PBG) structures, containing hundreds of nanometers of metals, which remain relatively transparent in a pre-established range, the visible light, and opaque for infrared and ultraviolet light [26]. This interesting property could be useful for many wide-ranging applications, from micro opto-electromechanical systems [27], or the enhancement of the quadratic and cubic nonlinearities of metal [28]. Propagation of light through the stack occurs as a result of a resonance tunneling phenomenon. This feature takes place for many transition metals ranging from group III to group XII. Some of the noble metals, such as Au, Ag or Cu, are relatively transparent within a transmission window located at the beginning of interband transition from the d band to conduction band [28]. The most important objective of a metallo-dielectric, multilayer structure is to enhance the transmitted field with respect to the transmission through an equivalent amount of metal in a certain range of wavelengths. In this way, one is able to use a large quantity of metal many skin depths thick while maintaining losses exceptionally small. In fact, the concept of skin depth (as the distance at which the field intensity is decreased to 1/e of its input value) looses its meaning due to the presence of spatial discontinuities that drastically modify the physical properties of the structure [26]. The bandwidth and the spectral position of the zero reflection condition, i.e. transparent metal regime, can be tuned by properly choosing the thickness of each layer. One or two anti-reflection coatings can be efficiently dimensioned by calculating the phase shift induced by the metal within the structure to improve the transmittance [29]. High values of transmittance and a high sensitivity to the presence of small quantity of the analyte, i.e. hydrogen, are reachable by using Pd, which is a transition metal, inside a periodic, metallo-dielectric structure. Several solutions for hydrogen sensing using Pd have been proposed by embedding a thin film of metal between an isolating film and the tip of an optical fiber [30], or by coating a piece of fiber with a sensitive material [31]. As suggested by the theory of metallo-dielectric multilayer structure, the overall thickness of Pd within the whole device can be increased, while keeping the device transparent to the visible light. As an example, we compare the transmittivity of a 50nm of pure Pd film grown on a glass substrate with that of a multilayer structure, composed of 4 periods and a half of Pd/dielectric (the dielectric material considered is ideal and has a relative permittivity equal to $\varepsilon_x=6.25$), embedded within two antireflection coatings. As depicted in Fig.2, we can notice an enhancement in the transmittivity by comparing the two

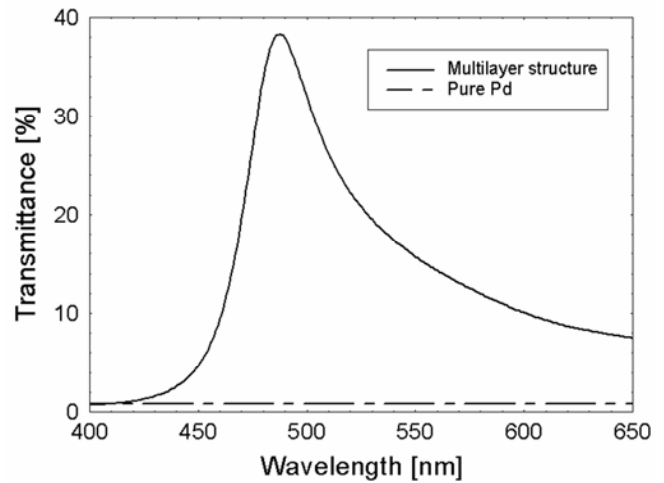

Fig.2: Comparison between the transmittivity of two structures containing the same quantity of Palladium: dashed line refers to a pure 50nm Pd film grown on a glass substrate; solid line refers to a multilayer metal-dielectric structure, composed of 4 periods and a half made by 10 nm of Pd and dielectric layers 85nm thick ($\varepsilon_x=6.25$). The peak value of T for the multilayer structure is 38,3% (487nm), compared with 0.78% of 50nm of pure Pd.

devices under test.

## III. OPERATING PRINCIPLE

As mentioned above, an efficient $H_2$-breath-test device has to detect small quantities of hydrogen in the breath exhaled, with high selectivity in the presence of other gases. For this reason traps for organics between the exhaled breath and the detecting device are useful in order to avoid damage or undesired chemical reactions. An efficient detection of hydrogen, operated by a multilayer structure, requires the use of an inert dielectric material that should be permeable to hydrogen. Furthermore, the optical properties of the dielectric material, i.e. the real and the imaginary part of the refractive index, must verify the following condition: the difference between the real part of the refractive index of the dielectric and the real part of the refractive index of palladium, in the visible range, must be sufficiently high to guarantee the resonance tunneling phenomenon. We start our analysis by considering a theoretical structure containing the sensitive material, Pd, and an ideal dielectric material X having $\varepsilon_x=6.25$ and an extinction coefficient equal to zero. Low values of hydrogen concentration, under the explosive limit of 4.65%, can be easily detected by the palladium/dielectric device if the optical properties of the hydride palladium are properly analyzed (Fig.1 a – b). If the surrounding mixture of gases is composed of oxygen (30%), nitrogen (70%) and hydrogen (a few ppm) only, we can extract information about the optical properties of the Pd films, in order to relate a pathologic condition with an optical response for the sensitive metal. The creation of a relation between the total hydrogen part in the mixture and the correspondent pathologic condition is very useful for our purposes. A malabsorption condition results in an increase from a baseline value of 20ppm in the exhaled breath. This value remains constant during the test for healthy people, and changes dramatically after few minutes for patients that suffer from lactose malabsorption or lactose intolerance. For medical purposes, we have to consider hydrogen values ranging from 20ppm to 100ppm, which are typical values for pathologic conditions [32]. Several works report on changes of the optical properties of thin films of hydride palladium absorbing different hydrogen concentrations, both under and over the explosive limit of hydrogen [21,22]. As reported above, through the evaluation of the nonlinear function *h* (see eq.1), all the parameters of the Drude model and the dielectric constant are calculated for all the hydrogen concentrations under investigation. The use of the multilayer, metallo-dielectric structure increases the total amount of sensitive material, i.e. palladium, and at the same time it improves the transmissivity. The multilayer structure under investigation is composed of five layers of 10nm of Pd alternated by 4 layers of 85nm of the X dielectric material. This structure is embedded within two antireflection coatings of 45 nm of the same dielectric X, as depicted in Fig.3. The antireflection layers thickness is calculated by considering the phase shift due to the presence of the metal within the structure for an operating wavelength equal to 550 nm. The transmission spectrum of this structure, reported in Fig.4, has been calculated using the standard transfer matrix method. Normal incidence, absorption and dispersion phenomena are considered in all the simulations. We emphasize that the dielectric layers are about eight times thicker than the palladium ones. By properly varying the dielectric thickness inside the multilayer structure one is able to narrow the resonance bandwidth, achieving values of full-width at half maximum of ~15nm for dielectric thicknesses of 500nm or more. We notice that this phenomenon does not modify the transmission properties of the structure, altering only the spectral response of the structure in terms of resonance spectral position and bandwidth. The transmission spectrum of the multilayer system undergoes small changes as the $H_2$ concentration is varied in the environment. For example, by increasing the concentration of hydrogen, increased transmittivity levels within the visible range is increased. This phenomenon is strictly linked to the red shift of the plasma frequency and to the enhancement of the damping factor of the palladium hydride as well. As depicted in Fig.5, we predict an increase of 6,3% in terms of peak transmission value when a malabsorption condition is detected, i.e. the overtaking of the 20ppm baseline. Moreover, a total increase of about 12,2% of the transmission spectrum is revealed when the total amount of hydrogen in the environment grows from 20ppm to 100ppm.

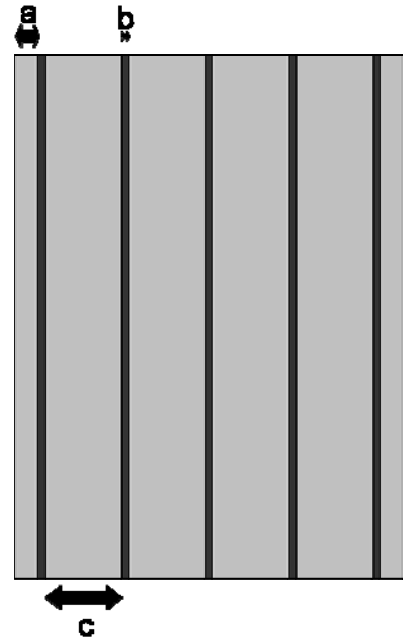

Fig.3: Metallo-dielectric sensor: light grey regions are the dielectric layers (X) and dark grey regions are the metal layers (Pd). The thicknesses of each layer are defined as a=45nm, c=85nm, b=10nm.

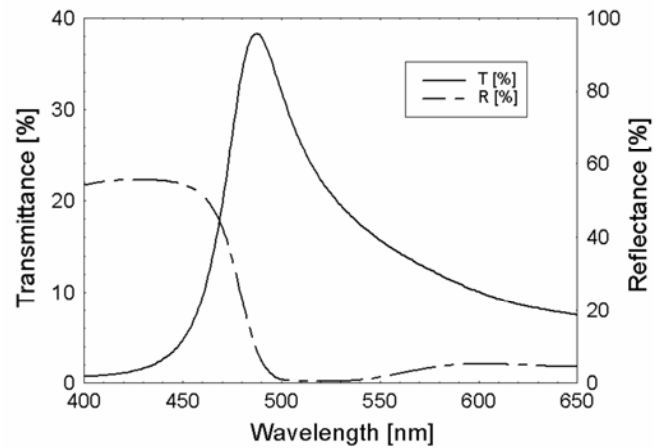

Fig.4: Sketch of transmission and reflection spectra for the multilayer structure.

## IV. Device Modeling

An efficient detection of hydrogen, operated by the multilayer structure, can be realized if a dielectric material inert and permeable to the hydrogen presence is used. Moreover, one of the most important issues to reach the transparent metal regime is ensured by a sufficiently high difference between the real part of the refractive index of the dielectric and the real part of the refractive index of palladium across the visible range. Polymers are the best candidates for our purposes, combining two of the most important requirements mentioned above: they are inert in the presence of hydrogen and their chemical features allow to be permeable to hydrogen molecules. While the majority of commercial polymers are characterized by a relatively low real part of the refractive index, reaching values lower than 1.8 [33], a high refractive index material has been recently proposed and realized. This material, known as EXP0454, is a titanium dioxide-rich polymer with a refractive index that reaches values close to 2 across the visible range. It shows high transparency and resistance to chemical attack [34]. This polymer is suitable for the realization of an $H_2$-breath-test device by all means. In fact, by properly alternating EXP0454 and Pd should be able to exploit the properties of a metallo-dielectric structure in order to realize a high sensitivity hydrogen sensor. As seen for the structure sketched in Fig. 3, the optimization of the transmission spectrum and spectral peak position depends on the parameters a, b and c. Another significant feature of the proposed device is the sensitivity of the structure against the variation of hydrogen concentration in the human exhaled breath. As shown in Fig. 5, the rise of the transmittance peak value against the hydrogen concentration could not be described exactly by a linear function. Our aim is to find a trade-off between an excellent sensitivity in the whole range of hydrogen concentration under investigation and the compactness of the device, obtained through the analysis of the transmission properties of the multilayer structure. As depicted in Fig. 6 an analysis of the sensor sensitivity has been performed by varying the parameter c (see Fig. 3). The sensitivity of the sensor is defined by the following expression:

$$S = \int_{H_2[ppm]} \frac{\Delta T_{peak}[\%]}{\Delta H_2[ppm]}, \quad H_2[ppm] = 20,...100 \text{ ppm} \quad (3)$$

We have performed an optimization analysis of the key parameters in order to find a balance between the width of the whole structure, the transmission features, including the transmission peak bandwidth, and the device sensitivity in the whole range of hydrogen concentration under investigation. According to Fig. 6 it is difficult to describe the sensitivity value (see eq.3) as a regular function. The maximum of sensitivity is reached at c=110nm (Fig.6). As depicted in

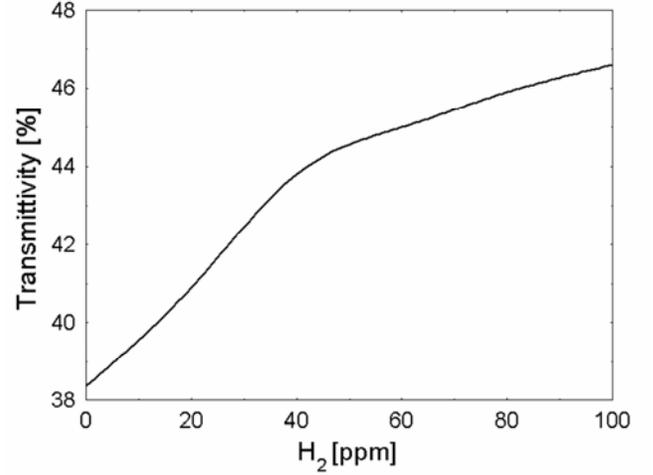

Fig.5: Metallo-dielectric sensor calibration curve: the transmission spectrum peak value (487 nm) increases by growing the concentration of

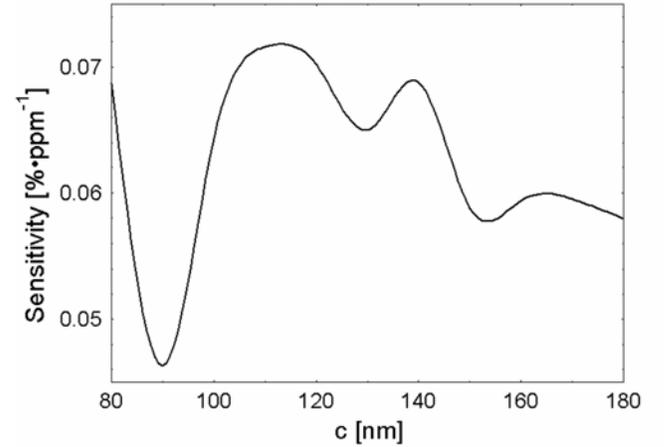

Fig6: Hydrogen sensor sensitivity (S) as a function of dielectric layer thickness, defined as c

Fig.7, the calibration curve for c=110nm shows the highest average sensitivity S, but it has a flat shape in the 40÷60ppm range. In this interval the device is not sufficiently sensitive to small variations of the percentage of $H_2$, and it does not appear to be useful for an $H_2$-breath-test. Looking for a better configuration, able to ensure good sensitivity in the whole range of investigation, we have analyzed configurations

characterized by other values of S (Fig.6). Among these, we found that c=120nm represents the best compromise in terms of sensitivity, device compactness and linearity of the calibration curve (Fig.7). In Tab.1 we summarized some features for the devices characterized by c=110nm and c=120nm. In particular we report on sensitivity of the devices (S) their total width (W) and transmission peak full-width at half maximum (B). No considerable variations of sensitivity are found by varying the total thickness of palladium layers in the structure (parameter b in Fig.3). The transmission spectrum of the optimized sensor (in this context, it is clear that we have not exhausted all of parameter space) is depicted in Fig.8. This spectrum is produced by a multilayer structure similar to the one sketched in Fig.3, using the following values: a=65nm, b=10nm, c=120nm. The measured transmittance allows us to plot the curve that relates the exhaled hydrogen concentration during the test after the ingestion of the lactose solution. After the ingestion the hydrogen levels in the breath are measured at regular intervals of 30 or 40 minutes.

TABLE I

| c [nm] | S [%·ppm$^{-1}$] | W [μm] | B [nm] |
|---|---|---|---|
| 110 | 0.0716 | 0.596 | 53 |
| 120 | 0.0702 | 0.636 | 61 |

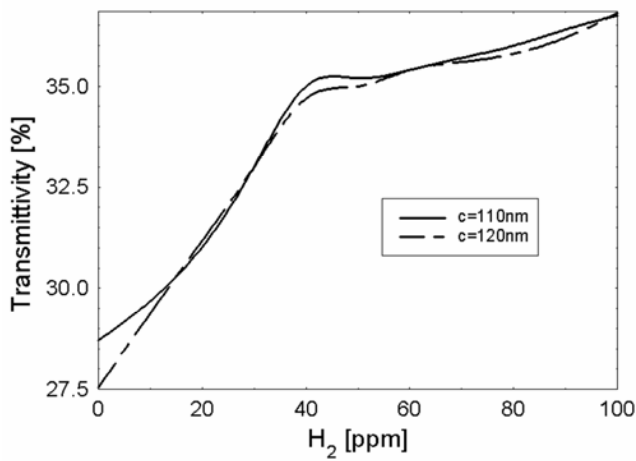

Fig.7: Metallo-dielectric sensor calibration curves for different dielectric thicknesses: c=110 nm (solid line) and c=120 nm (dashed line). Even if the average sensitivity value is approximately equal for these structures the one corresponding to c=120 nm is more appropriate to the breath test analysis, showing a gradual increase of the transmission peak value especially between 40 and 60ppm. The transmission peak position occurs at 508nm (solid line) and at 544 nm (dashed line).

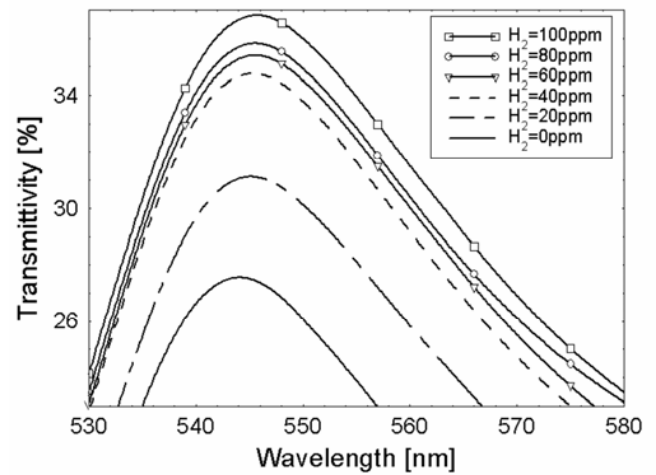

Fig.8: Transmission spectra for the optimized $H_2$ sensor : an increase of 11.6% in terms of transmittivity is noticed when a malabsorption condition occurs, i.e. the overtaking of the 20ppm baseline; the total increase of the transmittivity swelling the hydrogen concentration from 20ppm to 100ppm is about 16%.

V. CONCLUSION

We have proposed a novel approach to hydrogen sensing in terms of a compact, nano-optical sensor suitable for $H_2$-breath-test analysis. This kind of device is potentially very useful for the detection of lactose malabsorption or intolerance condition, and provides clear information about the healthy-unhealthy threshold. Moreover, this structure can be used to reveal the amount of hydrogen in the breath exhaled by unhealthy people, and can be used to establish a relationship between time and exhaled hydrogen, resulting in an efficient $H_2$-breath-test analysis, lactose malabsorption diagnosis in particular.

**Nicola Cioffi** is graduated in Chemistry -*summa cum laude*- at the University of Bari in 1997. In 2001 he received his *Ph.D.* in *Chemistry of Innovative Materials* at the University of Bari, working on the development, characterisation and analytical application of new nanostructured materials such as metal nanoparticles, organic/inorganic nano-composites and conductive polymer thin films. In 2001 he received the "*Best Young Researcher*" Award from the Analytical Chemistry Division of the Italian Chemical Society (SCI). From 2001 to 2005 he had several post-doc fellowships, still working on innovative nano-materials. Since April 2005, he is a researcher at the Chemistry Department of the University of Bari.

He has been member of the Organizing and/or Scientific Committee of national and international conferences; moreover, he his an active member of many Scientific Societies. In particular, since 2003, he is member of the *Analytical Spectroscopy Board* of the SCI - Analytical Chemistry Division.

He serves as referee for several journals and is *Grant Reviewer* for the *Health, Welfare and Food Bureau* of the Hong Kong Government.

*The scientific production of Nicola Cioffi* is documented by 50 papers on peer-reviewed journals and textbooks, and 77 communications presented in national and international conferences. He is also author of a *state-of-the-art* review chapter dealing with metal-fluoropolymer vapour sensors, published on the *Encyclopedia of Sensors* (details at www.aspbs.com/eos).

**Domenico de Ceglia** was born in Terlizzi (Bari-Italy) in 1977 and he graduated in Electronic Engineering at the Politecnico di Bari, in July 2003. He is currently enrolled in the Ph.D. program in Electronic Engineering. His main research activities concern the study of linear and nonlinear effects in photonic structures, waveguides, negative index materials, and the analysis of pulse propagation in nonlinear devices. He is currently a guest researcher at the Charles M. Bowden Research Center, Redstone Arsenal, Huntsville (Alabama-USA).

**Marco De Sario** was born in Cellamare (Bari-Italy) in 1942. He graduated in Electrical Engineering at the University of Bari in 1968. Since 1986 he is a full professor in Electromagnetic Fields. His main research interests are in microwave amplifiers, micro-strip devices, optical fibers, integrated optics, optical non-linear devices, optical amplifiers and laser in chalcogenide glasses, and photonic band-gap devices. Prof. De Sario is author or co-author of about 250 scientific papers published on prestigious, high impact factor international magazines, congresses, conferences, etc. Marco De Sario is a referee of many scientific international magazines as IEEE Journals, OSA Journals, etc., and referee of the Politecnico of Bari in the Italian Electromagnetic board.

**Antonella D'Orazio** was born in Potenza (Italy) in 1958. She received the Dr. Eng. degree, summa cum laude, in Electrical Engineering at the University of Bari in 1983 and the Ph.D. in Electromagnetisms in 1987. Since 1983 she has joined the Dipartimento di Elettrotecnica ed Elettronica of the Politecnico di Bari. Presently she is full professor in Electromagnetic Fields. Since 2003 she is Coordinator of the Teacher Council of the Phylosophiae Doctorate in Electronics Engineering. Her main research interests are the design, fabrication and characterization of integrated optic, linear and nonlinear devices based on titanium diffused lithium niobate waveguides and ion-exchanged waveguides, the numerical modeling of photonic band gap devices. She has co-authored over 200 publications published on international journals and conference proceedings, lectures and invited papers. She is involved in several national and international projects and co-operations. She is currently referee of a number of international journals. She served as member of the program committee for many national and international workshops and conferences. Since 2003 she is involved in the activities of the COST P11 action "Physics of linear, nonlinear and active phoyonic crystals" and, also as member of the Management Committee appointed by the Italian Ministry of University and Scientific Research, in the activities of the COST 288 action "Nanoscale and ultrafast photonics". Prof. D'Orazio is a member of the Federazione Italiana di Elettrotecnica, Elettronica, Automazione, Informatica e Telecomunicazioni (AEIT), of the Institute of Electrical and Electronics Engineers (IEEE) and of the Optical Society of America (OSA).

**Vincenzo Petruzzelli** was born in Bari (Italy) in 1955 and graduated in Electrical Engineering at the University of Bari in 1986. Since then he has joined the Dipartimento di Elettrotecnica ed Elettronica of Politecnico di Bari, receiving in 1992 the Ph. D. in Electronic Engineering. He is currently associate professor of Politecnico di Bari. Now he is engaged in theoretical and technological research on integrated optic devices, titanium diffused lithium niobate waveguides, ion-exchanged waveguides, nonlinear optics, active devices. He is co-authored about 160 journal and conference papers.

**Francesco Prudenzano** was born in Manduria (Taranto-Italy) in 1964. He received the Degree in Electronic Engineering at University of Bari in 1990 and the PhD Degree in Electronic Engineering in 1996. Since 1990 he has been with the Dipartimento di Elettrotecnica ed Elettronica of the Politecnico of Bari where he is currently associate professor. His research work includes the design of integrated optical linear and nonlinear, active devices, numerical modeling of photonic and electromagnetic band gap structures and microstructured fibers.

**Michael Scalora** was born in Solarino, (Siracusa-Italy), on November 9, 1962. He earned a BS degree in Physics from Montclair State University, Montclair, NJ, in 1984, a Master of Science and a PhD in Physics from Rensselaer Polytechnic Institute, Troy, NY, in 1988 and in 1990, respectively. He is currently with the Charles M. Bowden Research Center, Redstone Arsenal, Alabama.

**Simona Trevisi** was born in Bari, Italy, in 1983 and she is graduating in Electronic Engineering at Politecnico di Bari.

**Maria Antonietta Vincenti** was born in Bari, Italy, in 1982 and graduated in Electronic Engineering, summa cum laude, at the Politecnico di Bari in December 2005. In May 2006 she joined the Dipartimento di Elettrotecnica ed Elettronica of Politecnico di Bari, as a Ph.D. student in Electronic Engineering. Her main research activities concern the study of optical sensors for medical applications and imaging techniques based on metallo-dielectric structures.